\documentclass[aps,prd,twocolumn,superscriptaddress,floatfix,nofootinbib]{revtex4-1}

\usepackage{amscd}
\usepackage{amssymb}
\usepackage{amsmath}
\usepackage{mathtools}
\usepackage{multirow}

\usepackage[colorlinks=true]{hyperref}
\usepackage{graphicx}
\graphicspath{{./plots/}}
\usepackage[sort&compress]{natbib}
\bibpunct{[}{]}{,}{n}{}{}

\usepackage{usebib}
\newbibfield{author}
\newbibfield{journal}
\newbibfield{volume}
\newbibfield{pages}
\newbibfield{year}
\newbibfield{eprint}
\bibinput{pcm_continuity}

\RequirePackage{pdfcomment}
\RequirePackage{etoolbox}

\newcommand{\ubauthortitle}[1]{\usebibentry{#1}{author}, ``\usebibentry{#1}{title}''}
\newcommand{\ubjref}[1]{\usebibentry{#1}{journal} \usebibentry{#1}{volume} (\usebibentry{#1}{year}) \usebibentry{#1}{pages}}
\newcommand{\ubeprint}[1]{[ArXiv:\usebibentry{#1}{eprint}]}
\newcommand{\ubref}[1]{\ubauthortitle{#1}, \ubjref{#1}, \ubeprint{#1}}

\let\oldcite\cite

\renewcommand{\cite}[1]{\oldcite{#1}%
\renewcommand{\do}[1]{ + \ubref{##1}\textCR}%
\pdftooltip{${}^{>}$}{\docsvlist{#1}}%
}

\newcommand{\xddots}{%
  \raise 4pt \hbox {.}
  \mkern 6mu
  \raise 1pt \hbox {.}
  \mkern 6mu
  \raise -2pt \hbox {.}
}

\newcommand{\comment}[1]{}

\newcommand{\lr}[1]{ \left( #1 \right) }
\newcommand{\lrs}[1]{ \left[ #1 \right] }

\newcommand{\Tr}{ {\rm Tr} \, }
\newcommand{\tr}{ {\rm tr} \, }
\newcommand{\re}{ {\rm Re} \, }

\def \kv{\mathbf{k}}

\def \xv{\mathbf{x}}
\def \yv{\mathbf{y}}
\def \ev{\mathbf{e}}

\def \Tr{\mathrm{Tr}}
\def \tr{\mathrm{tr}}

\def \exp{\mathrm{exp}}

\def \ipr{\mathrm{IPR}}

\begin{document}
\sloppy

\title{Lattice study of continuity and finite-temperature transition in two-dimensional \texorpdfstring{$SU(N) \times SU(N)$}{SU(N) x SU(N)} Principal Chiral Model}
\thanks{This paper includes PDF tooltips for citations. Please hover the mouse over the $>$ symbol after citations to see a tooltip with bibliographic information.}

\author{P.~V.~Buividovich}
\email{pavel.buividovich@physik.uni-regensburg.de}
\affiliation{Institute for Theoretical Physics, Regensburg University, D-93053 Regensburg, Germany}

\author{S.~N.~Valgushev}
\email{semen.valgushev@ur.de}
\affiliation{Institute for Theoretical Physics, Regensburg University, D-93053 Regensburg, Germany}
\affiliation{Institute for Theoretical and Experimental Physics (ITEP), 117218 Russia, Moscow, B. Cheremushkinskaya str. 25}

\date{June 27th, 2017}
\begin{abstract}
We present first-principle lattice study of the two-dimensional $SU(N) \times SU(N)$ Principal Chiral Model (PCM) on the cylinder $\mathbb{R} \times S^1$ with variable compactification length $L_0$ of $S^1$ and with both periodic and $Z_N$-symmetric twisted boundary conditions. For both boundary conditions our numerical results can be interpreted as signatures of a weak crossover or phase transition between the regimes of small and large $L_0$. In particular, at small $L_0$ thermodynamic quantities exhibit nontrivial dependence on $L_0$, and the static correlation length exhibits a weak enhancement at some ``critical'' value of $L_0$. We also observe important differences between the two boundary conditions, which indicate that the transition scenario is more likely in the periodic case than in the twisted one. In particular, the enhancement of correlation length for periodic boundary conditions becomes more pronounced at large $N$, and practically does not depend on $N$ for twisted boundary conditions. Using Gradient Flow we study non-perturbative content of the theory and find that the peaks in the correlation length appear when the length $L_0$ becomes comparable with the typical size of unitons, unstable saddle points of PCM. With twisted boundary conditions these saddle points become effectively stable and one-dimensional in the regime of small $N L_0$, whereas at large $N L_0$ they are very similar to the two-dimensional unitons with periodic boundary conditions. In the context of adiabatic continuity conjecture for PCM with twisted boundary conditions, our results suggest that while the effect of the compactification is clearly different for different boundary conditions, one still cannot exclude the possibility of a weak crossover separating the strong-coupling regime at large $N L_0$ and the Dunne-\"{U}nsal regime at small $N L_0$ with twisted boundary conditions.
\end{abstract}

\maketitle

\section{Introduction}
\label{sec:introduction}

The innovative idea of resurgent trans-series has recently allowed to better understand the structure of perturbative expansions for asymptotically free quantum field theories, such as four-dimensional gauge theories and two-dimensional sigma models. In particular, for two-dimensional sigma models resurgent trans-series provide a precise interpretation of the factorial non-Borel-summable infrared renormalon divergences of perturbative series \cite{Unsal:12:1,Dunne:12:1,Unsal:14:1,Cherman:14:1,Unsal:15:1} in terms of saddle points of the classical action, even if they are non-topological, unstable and/or complex-valued.

However, at present this interpretation of infrared renormalon divergences can be explicitly worked out only for quantum field theories with compactified spatial direction $\mathbb{R}^{d-1} \times S^1$, in which the fields satisfy certain twisted boundary conditions. The compactification length $L_0$ should be sufficiently small with $N L_0 \Lambda \ll 2 \pi$, where $\Lambda$ is the dynamically generated energy scale and $N$ is the rank of the symmetry group. In this limit the theory is in the weakly coupled regime while still exhibiting non-perturbative features such as dynamically generated mass gap. This defines the so-called \"{U}nsal-Dunne regime, which allows for explicit construction of trans-series.

The so-called continuity conjecture states that the \"{U}nsal-Dunne regime at small $L_0$ is analytically connected to the strongly coupled regime at large $L_0$ \cite{Unsal:14:1,Cherman:14:1}, in which boundary conditions become irrelevant and the physics is equivalent to the low-temperature phase with periodic boundary conditions. This conjecture is based on the observation that the physical properties of gauge theories and sigma models appear to be qualitatively very similar in both \"Unsal-Dunne regime and in the genuine strong-coupling regime at small temperatures or large compactification length.

The continuity conjecture is closely related to the Eguchi--Kawai (EK) reduction in lattice field theory \cite{Eguchi:82:1,Okawa:83:1,Okawa:10:1} where the full theory is suggested to be equivalent to twisted single site model in the large $N$ limit. It is known that the original EK reduction without twist \cite{Eguchi:82:1} does not work due to spontaneous breaking of the center symmetry $\mathbb{Z}_N^d$ \cite{Bhanot:82:1}. Considering EK reduced model as a result of continuous dimensional reduction from large to very small lattices it is evident that this symmetry breaking is manifestation of deconfinement phase transition happening when the torus is sufficiently small $L_0 \Lambda \sim 1$ \cite{Neuberger:03:1,Neuberger:03:2}. A possible solution is to introduce the twisted boundary conditions which preserve center symmetry and prevent it from spontaneous breaking, thus suppressing the deconfinement transition and allowing for analytic (or volume independent) connection between the regimes of small and large $L_0$. However, lattice simulations of twisted EK reduced model indicate that spontaneous symmetry breaking can still occur and pose the question of the existence of the continuum limit \cite{Azeyanagi:08:1}, although it seems that these difficulties can be overcome \cite{Okawa:10:1}. Possible ways of stabilizing center symmetry in gauge theories are the special deformations of the gauge action \cite{Unsal:08:1} or the introduction of adjoint fermions \cite{Unsal:07:1}, which effectively induce center-preserving holonomies along the compactified directions. These ideas were important to formulate the continuity conjecture in PCM.

In two-dimensional sigma models the prescription for $\mathbb{Z}_N$-preserving twist $\Omega \in SU(N)$ reads as:
\begin{eqnarray}
\label{eq:twist_prescription}
\Tr\Omega^n =
\begin{cases}
N, & n \equiv 0 \, \mathrm{mod} \, N \\
0, &\mbox{otherwise}
\end{cases}.
\end{eqnarray}
This operator either projects excited states out or provides a phase shift which leads to mutual cancellations between distinct states in the partition function. Since a lot of excited states do not contribute at all to the twisted partition function \cite{Sulejmanpasic:16:2}, one can hope that the deconfinement transition is eliminated \cite{Unsal:14:1,Cherman:14:1}. However, to turn continuity conjecture into a precise statement, one should demonstrate that no phase transition or crossover occur as the compactification length $L_0$ changes from large values $N L_0 \Lambda \gg 2 \pi$ to small values with $N L_0 \Lambda \ll 2 \pi$. At present a rigorous analytic demonstration of this fact is still lacking due to the absence of reliable analytic methods for strongly coupled quantum field theories. Notable exceptions are the exactly solvable large-$N$ $\mathbb{CP}^{N-1}$ and $O(N)$ non-linear sigma-models for which an explicit demonstration has been worked out \cite{Sulejmanpasic:16:2}. However, for Principal Chiral Model (PCM) which is especially interesting due to its matrix-like planar limit very similar to that of QCD the problem clearly calls for first-principle simulations.

Unfortunately, not much is known about thermodynamic properties of PCM in general, although this model is integrable and many exact results can be obtained using bootstrap techniques \cite{Orland:11:1,Orland:12:1,Orland:13:1,Orland:14:1,Orland:16:1}. One of the reasons is that for PCM there is no obvious local parameter which can be used to characterize the ``deconfinement'' phase transition, rendering analytic and lattice studies very difficult. Recently a thermodynamic Bethe ansatz has been proposed in order to investigate thermodynamic properties \cite{Cubero:15:1}, however without definite conclusions so far.

In this work we test continuity conjecture for the two-dimensional $SU(N) \times SU(N)$ PCM using first-principle Monte-Carlo simulations. We study several characteristic quantities such as static correlation length, mean energy and specific heat and demonstrate that they exhibit qualitatively different dependence on the length of the compact direction $L_0$ with periodic (Section~\ref{sec:pbc}) and twisted (Section~\ref{sec:tbc}) boundary conditions. In both cases we find some evidence for a transition/crossover which however posses very different features: while for periodic boundary conditions this might be a finite-temperature transition (probably similar to deconfinement in QCD), for twisted boundary conditions this is at most a crossover with respect to the combined length parameter $\rho \equiv N L_0$.

Furthermore, in Section~\ref{sec:gradient_flow} we use Gradient Flow \cite{Luscher:10:1} to evolve the field configurations sampled by Monte-Carlo process towards the saddle points of the classical action, and demonstrate that the resulting ``almost classical'' field configurations feature localized non-perturbative objects which resemble the uniton and fracton saddle points known for continuum PCM \cite{Unsal:14:1,Cherman:14:1}. Twisted boundary conditions stabilize those saddle points and, as expected, lead to the emergence of effective topological sectors \cite{Unsal:14:1,Cherman:14:1}. We also find that geometric properties of non-perturbative objects strongly change in the region of the possible phase transition or crossover for both boundary conditions.

\section{Simulation setup and observables}
\label{sec:setup}

The lattice action of the two-dimensional $SU(N)\times SU(N)$ PCM can be written as:
\begin{eqnarray}
\label{eq:lattice_action}
S\lrs{U\lr{\xv}} = -2\beta N \sum_{\xv,i} \re \Tr\lrs{U\lr{\xv}U^\dagger\lr{\xv + \ev_i}},
\end{eqnarray}
where $\beta \equiv \lambda^{-1} = 1/(g^2 N)$ is an inverse of the t'Hooft coupling $\lambda$ and $\ev_i$ is a unit lattice vector in direction $i$. We have used lattices of the size $L_0 \times L_1$ with boundary conditions (BC) defined as:
\begin{eqnarray}
\label{eq:boundary_conditions}
U\lr{x_0+L_0,x_1} &=&	 \Omega_0 U\lr{x_0,x_1} \Omega_0^\dagger, \nonumber \\
U\lr{x_0, x_1+L_1}&=&U\lr{x_0,x_1},
\end{eqnarray}
where matrix $\Omega_0$ determines the type of boundary conditions:
\begin{eqnarray}
\label{eq:bc_matrix}
\Omega_0=
\begin{cases}
I, & \mbox{for periodic BC (PBC)} \\
\Omega, & \mbox{for twisted BC (TBC)}
\end{cases} .
\end{eqnarray}
The twist matrix $\Omega$ has the following form:
\begin{eqnarray}
\label{eq:twist_matrix}
\Omega=e^{i\frac{\pi}{N}\nu} \mathrm{diag} \{ 1, e^{i \frac{2\pi}{N}}, \cdots , e^{i \frac{2\pi\lr{N-1}}{N}} \},
\end{eqnarray}
where $\nu=0,1$ for $N$ odd, even. It is easy to see that $\Omega$ satisfies the equation (\ref{eq:twist_prescription}).

We employed the standard Cabbibo-Marinari algorithm \cite{Cabibbo:82:1} in order to stochastically sample field configurations $U(\xv)$ according to Boltzmann weight $\exp\lr{-S\lrs{U\lr{\xv}}}$ with the action (\ref{eq:lattice_action}). One Monte-Carlo update of the field configuration was implemented by applying the heat bath algorithm to all $SU(2)$ subgroups of all $U(\xv)$ matrices. Each Monte-Carlo step was followed by overrelaxation step \cite{Adler:81:1} in order to decrease auto-correlation time. Error analysis was carried out using jacknife and bootstrap techniques. In order to test our code, we have reproduced several data points from the previous PCM simulations of \cite{Rossi:94:1} with very high precision.

We performed calculations for $\beta = 0.332$, $N=6,9,12,18$ and spatial lattice sizes $L_1 = 108$ and, for $N = 18$, $L_1 = 200$, which we found to be sufficiently large compared to zero-temperature static correlation length taking values in the range $\xi_0 = 10 \ldots 12$ (for different $N$) in our simulations. For each value of $N$ and type of boundary conditions we have simulated at multiple values of the compactification length $L_0$ in the intervals $1 \leq L_0 \leq L_1$ for periodic and twisted boundary conditions, respectively.

In order to study basic thermodynamic properties, we have computed the mean energy
\begin{eqnarray}
\label{eq:energy_density}
E = 1 - \frac{1}{4 N^2}\frac{\partial F}{\partial \beta} = \nonumber \\
=1 - \frac{1}{N} \bigg \langle \re\Tr\lrs{U\lr{\xv}U^\dagger\lr{\xv+\ev_i}} \bigg \rangle
\end{eqnarray}
and specific heat
\begin{eqnarray}
\label{eq:specific_heat}
C = \frac{1}{N} \frac{d E}{d g^2} .
\end{eqnarray}
In order to compute the static correlation length we rely on the observation that in the weak-coupling regime for sufficiently small values of lattice momenta as compared to the dynamically generated mass gap the Fourier transform of the group invariant correlation function $G(\xv)$ can be with a good precision approximated by the free scalar propagator \cite{Rossi:94:1}:
\begin{eqnarray}
\label{eq:correlatior_free_bosons}
\tilde G(\kv) \simeq \frac{Z_G}{M^2 + 4 \sin^2\lr{k_0/2} + 4 \sin^2\lr{k_1/2}},
\end{eqnarray}
where $Z_G$ is the wave-function renormalization constant. From this equation one obtains an expression for static correlation length $\xi$ in terms of the momentum-space correlator at the two lowest values of lattice momenta:
\begin{eqnarray}
\label{eq:correlation_length_Rossi}
\xi^2 = \frac{1}{4\sin^2 \lr{\pi/L}}\lrs{\frac{\tilde G(0,0)}{\tilde G(0,1)}-1}.
\end{eqnarray}
Alternatively, correlation length can be computed with the standard exponential fits of wall-wall correlators and we have checked that numerical values for $\xi$ obtained in both ways agree with a high precision.

Similarly to the two-dimensional lattice gauge theory \cite{Gross:80:1}, lattice PCM undergoes the large $N$ phase transition from strong to weak coupling phase at $\beta_c = 0.305$ \cite{Vicari:94:1,Rossi:94:1,Rossi:94:2}, which in this case is a second-order transition at which the specific heat diverges and the distribution of eigenvalues $\lambda = e^{i\phi}$ of the link matrices $U\lr{\xv} U^\dagger\lr{\xv+\ev_i}$ develops a gap. Continuum scaling of different observables such as mass gap already sets up in the vicinity of the critical coupling $\beta \gtrsim \beta_c$ \cite{Vicari:94:1,Rossi:94:1}. For our simulations we have chosen the value $\beta = 0.332$ sufficiently deep in the weak coupling phase. To check that with this value of coupling we are sufficiently to the continuum limit, we have observed the asymptotic scaling of zero-temperature static correlation length $\xi_0$ (measured on lattices with $L_0 = L_1$ for $N \geq 9$) in the so-called $\beta_E$ scheme \cite{Vicari:94:1}:
\begin{eqnarray}
\label{eq:continuum_scaling}
\xi(E) = \frac{0.991}{16\pi} e^{\frac{2-\pi}{4}} \sqrt{E} e^{\pi/E} .
\end{eqnarray}
While the constraints in CPU time available for our simulations have forced us to work at a single value of $\beta$, in future work it would be also desirable to check the continuum scaling of correlation length and thermodynamic observables by performing simulations at several values of $\beta$.

In order to exclude possible large $N$ phase transition to physically irrelevant strong coupling phase as the compactification length varies in the range $L_0 = 1 \dots L_1$, we have studied angle distribution of eigenvalues of link matrices $U\lr{\xv} U^\dagger\lr{\xv+\ev_i}$ and found that it remains gapped and almost unchanged for all directions $i$ and boundary conditions at all values of $L_0$. This suggests that finite temperature transition which we discuss below has different nature and seems to be a feature of continuum theory.

In order to better visualize important properties of our physical observables we have normalized them with respect to the corresponding zero temperature values at periodic boundary conditions and the same $N$:
\begin{eqnarray}
\label{eq:observable_0}
O_0 \equiv O(L_0 = L_1, \mathrm{PBC}),
\end{eqnarray}
where $O$ is some physical observable, and depict the relative change rather than the value itself:
\begin{eqnarray}
\label{eq:observable_relative_change}
\frac{\Delta O(L_0)}{O_0} \equiv \frac{O(L_0) - O_0}{O_0}.
\end{eqnarray}
This normalization is motivated by the fact that all observables admit non-trivial $1/N$ corrections which render straightforward comparison at different $N$ not very illustrative. Note also that identical normalization allows to conveniently compare the physics for both types of boundary conditions. We have also performed fits to the lowest-order $1/N$ expansion:
\begin{eqnarray}
\label{eq:observable_N_fit}
O(L_0,N) = \tilde O(L_0) + c_1 / N^2
\end{eqnarray}
in order to estimate infinite $N$ values $\tilde O(L_0)$, which are also presented on our figures. Very high precision of the numerical data for the mean energy also allowed us to include the terms of order $1/N^4$ into the fitting function (\ref{eq:observable_N_fit}) for this observable. In order to calculate the relative changes (\ref{eq:observable_relative_change}) in the limit $N \rightarrow \infty$, we perform separate large-$N$ extrapolations of the form (\ref{eq:observable_N_fit}) both for $O_0 \equiv O(L_0 = L_1, \mathrm{PBC})$ in the denominator and for the finite-temperature value $O(L_0)$ in the numerator, which typically results in a smoother extrapolation. Furthermore, when plotting the data points as functions of the compactification length $L_0$, we express $L_0$ in units of zero-temperature static correlation length $\xi_0$ calculated with the same $N$ as the data. This rescaling should allow to compare our current data with prospective simulations at other values of coupling as well as with calculations for continuum PCM.

\section{Periodic boundary conditions and the finite-temperature ``deconfinement'' transition}
\label{sec:pbc}

\begin{figure*}[h!tpb]
\centering
 \includegraphics[scale=0.8]{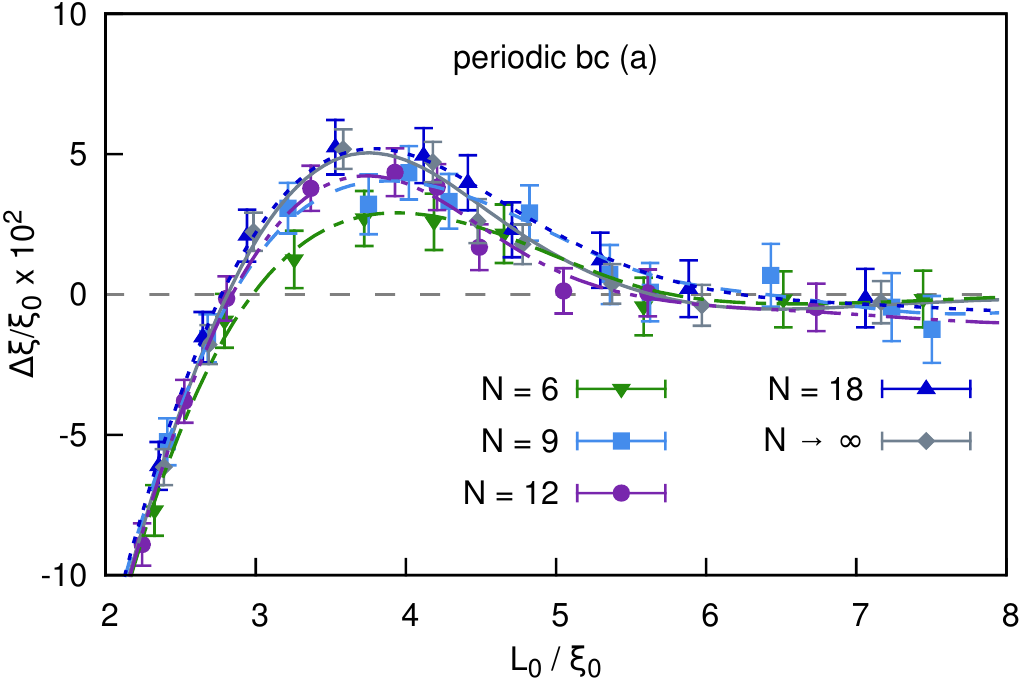}
 \includegraphics[scale=0.8]{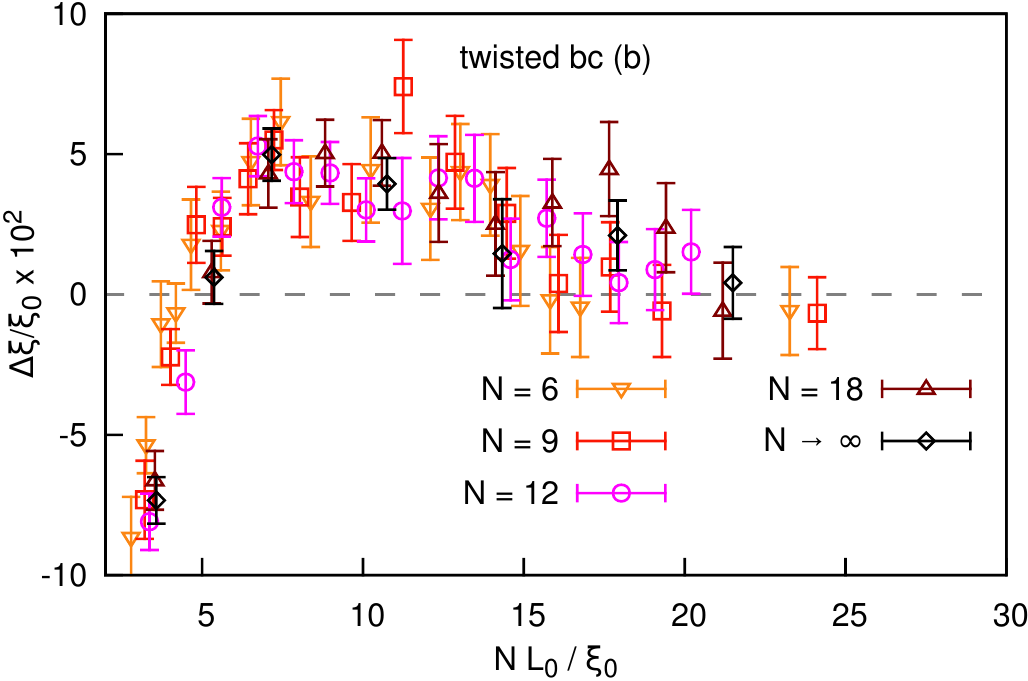} \\
 \includegraphics[scale=0.8]{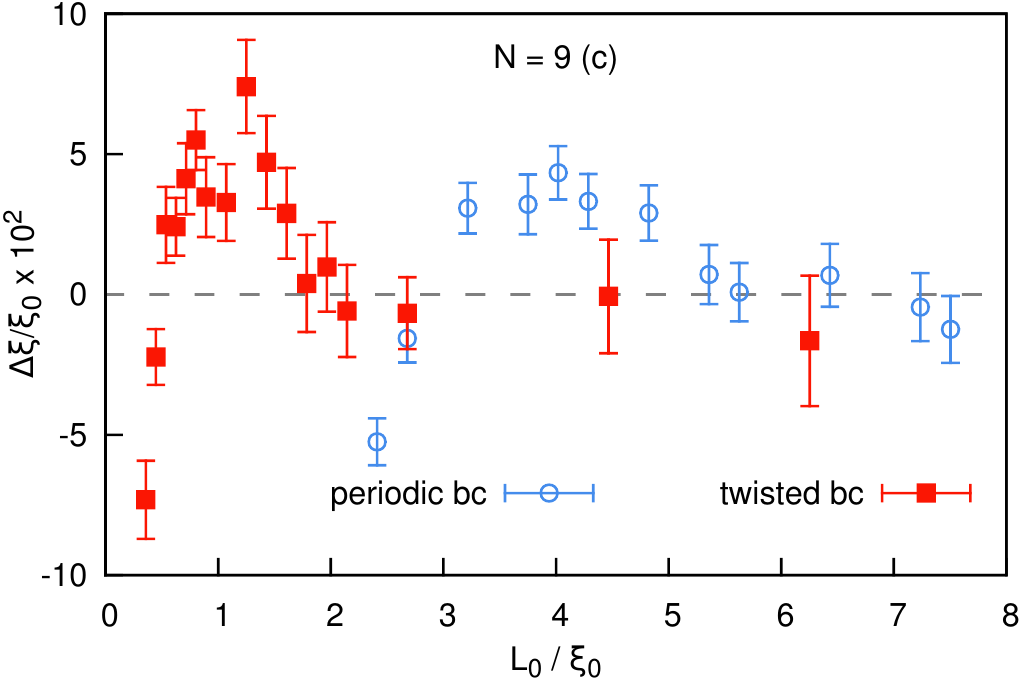}
 \includegraphics[scale=0.8]{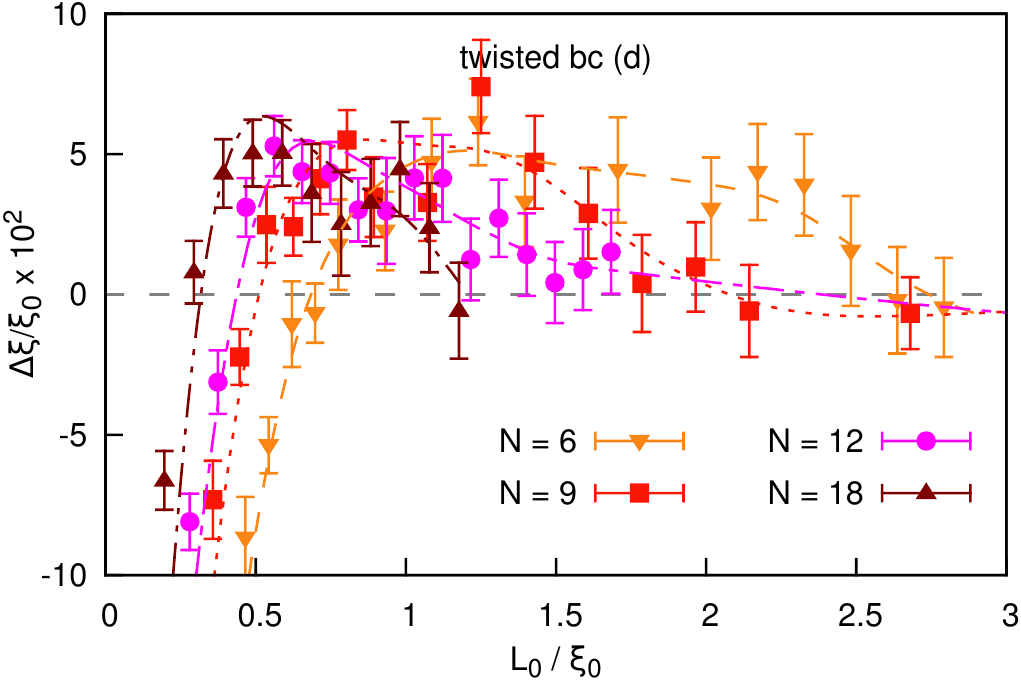} \\
 \caption{Relative change of correlation length $\Delta \xi/ \xi_0$ as a function of compactification length $L_0$ for different boundary conditions and values of $N$. In the plots at the top, we illustrate the dependence of the correlation length on the natural compactification scales: $L_0$ for PBC and $N L_0$ for TBC. The plots at the bottom illustrate how the peak in the correlation length with TBC shifts to smaller $L_0$ as $N$ is increased. For periodic boundary conditions, extrapolations to infinite $N$ are obtained using the fits of the form (\ref{eq:observable_N_fit}).}
\label{fig:correlation_length}
\end{figure*}

To study possible transition between low-temperature and high-temperature regimes, which correspond to large and small compactification length $L_0$ with periodic boundary conditions, on Fig.~\ref{fig:correlation_length}(a) we first illustrate the dependence of the static correlation length $\xi$ on $L_0$. At all values of $N$ the correlation length stays almost constant for large values of $L_0$, then exhibits statistically significant growth in the range $L_0/\xi_0 = 3 \dots 5$, and finally decreases for smaller $L_0$, seemingly reaching some finite value at $L_0 \rightarrow 0$. The position of the maximum of correlation length slowly shifts to smaller values when $N$ is increased. For the data points extrapolated to infinite $N$ the peak lies approximately at $L_0^c/\xi_0 \approx 3.5$. An important point to stress that in contrast to e.g. $CP^N$ sigma model, for the principal chiral model the transition temperature remains finite in the large-$N$ limit, as also indicated by the recent Diagrammatic Monte-Carlo simulations directly in the large-$N$ limit \cite{Buividovich:17:3}. This property is also expected for large-$N$ gauge theories \cite{Lucini:05:1}.

\begin{figure*}[h!tpb]
\centering
 \includegraphics[scale=0.82]{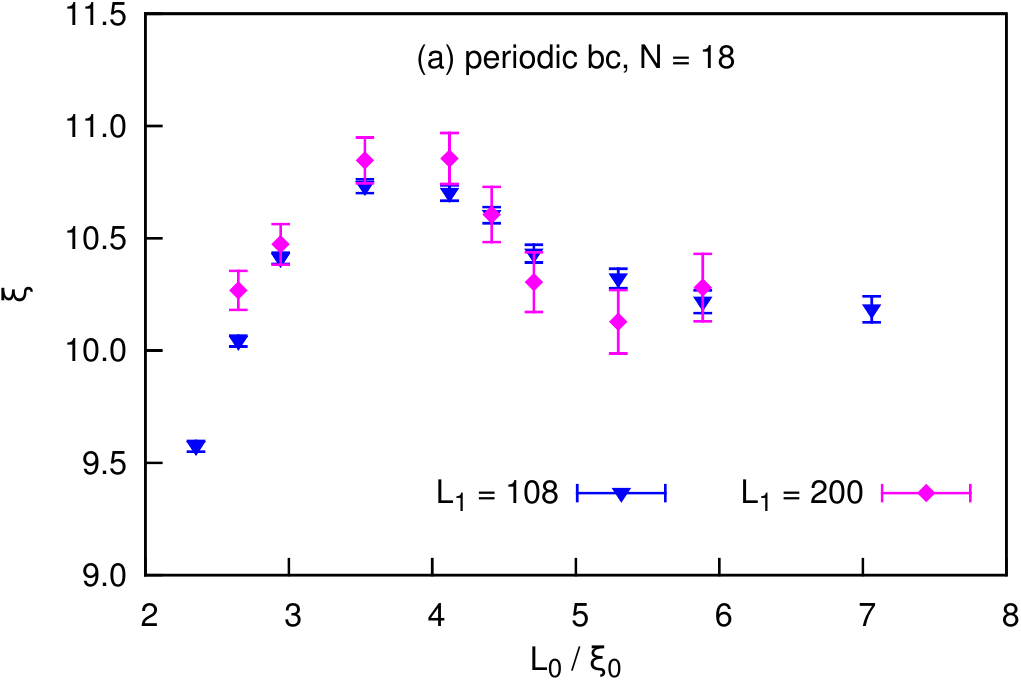}
 \includegraphics[scale=0.82]{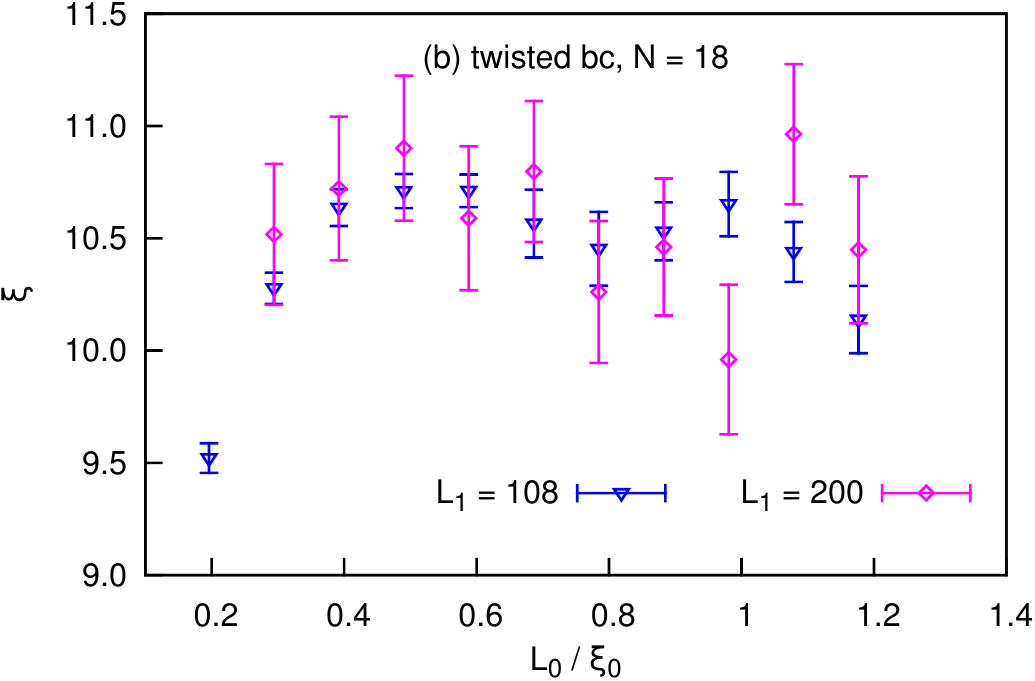}
 \caption{ Volume dependence of the correlation length $\xi$ in the vicinity of the peak for periodic and twisted boundary conditions for $N = 18$.}
\label{fig:corr_length_volume_scaling}
\end{figure*}

At the same time, the maximal value of the peak shows tendency to grow with $N$ and its width seems to decrease. Extrapolations to the large-$N$ limit using the fits of the form (\ref{eq:observable_N_fit}) at fixed values of $L_0$ suggest that the maximal value of $\xi$ remains finite in the large $N$ limit. Furthermore, even upon the large-$N$ extrapolation the enhancement of the correlation length appears to be very weakly pronounced: the relative change in $\xi$ is of order of $5\%$. Simulations at larger volumes ($L_1 = 200$) also reveal a rather small ($\sim 2 \%$) enhancement of $\xi$ with volume, see the left plot on Fig.~\ref{fig:corr_length_volume_scaling}.

\begin{figure*}[h!tpb]
\centering
 \includegraphics[scale=0.8]{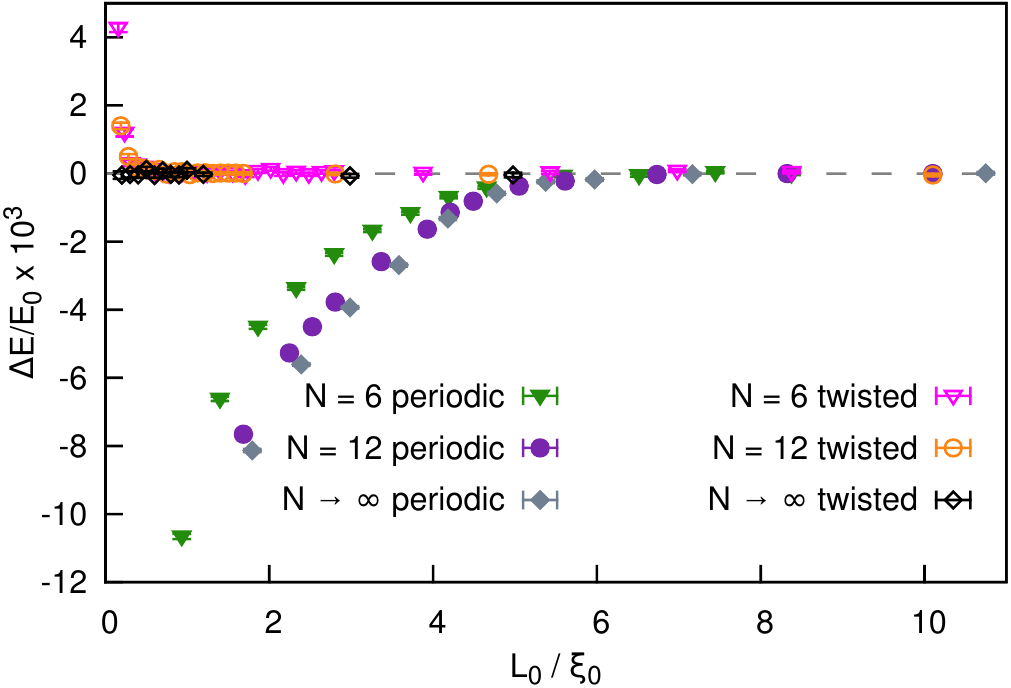}\includegraphics[scale=0.8]{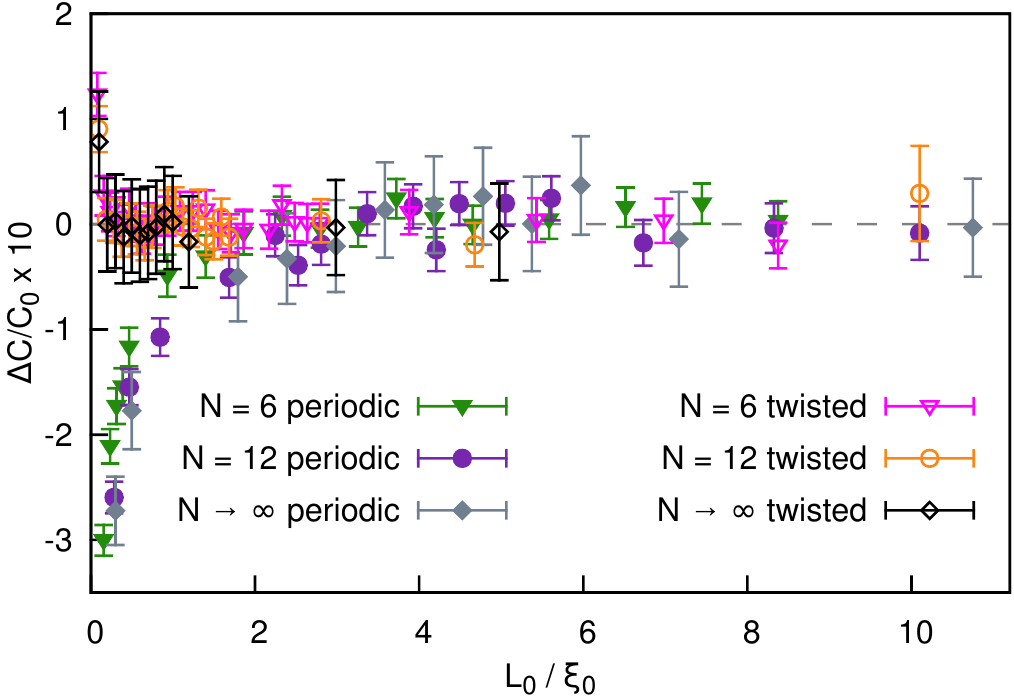}
 \caption{Relative changes of the mean energy $\Delta E / E_0$ (left plot) and the specific heat $\Delta C / C_0$ (right plot) as functions of compactification length $L_0$ for different values of $N$ and boundary conditions. The $N \rightarrow \infty$ data for the mean energy are obtained using extrapolations of the form $E(N) = \tilde E + c_1 / N^2 + c_2 / N^4$ at fixed $L_0$.}
\label{fig:mean_energy_and_specific_heat}
\end{figure*}

The emergence of such a peak structure and clearly distinct behaviour of correlation length at small and large $L_0$ are suggestive of a finite-temperature phase transition or crossover. This transition also manifests itself in the distinct behaviour of thermodynamic observables at low and high temperatures. In particular, both the mean energy $E(L_0)$ and the specific heat $C$ take almost constant values when $L_0 > L_0^c$ and then decrease in the region $L_0 < L_0^c$, see Fig.~\ref{fig:mean_energy_and_specific_heat} for illustration.

However, not much is known on the nature of this transition. The observed scaling with $N$ and $L_1$ suggest that the finite-temperature transition is rather weak, but cannot completely distinguish between the weak phase transition and crossover. A more detailed combined study of the finite-volume and finite-$N$ scaling is required to make a definite conclusion about the order of this finite-temperature transition, which we leave for future work. A recent Diagrammatic Monte-Carlo study \cite{Buividovich:17:3} at $N \rightarrow \infty$ limit also indicated the relatively weak enhancement of correlation length at the same critical value of $L_0$, but did not completely exclude the possibility of divergent correlation length at the transition point.

On the general grounds in analogy with other asymptotically free theories one might expect a ``deconfinement'' phase transition associated with effective liberation of $SU(N)$ degrees of freedom at sufficiently high temperature. For gauge theories, the deconfinement transition is typically associated with the breaking of the global $Z_N^d$ center symmetry, with Polyakov loop being the local order parameter. In contrast, for principal chiral model even an approximate local order parameter which would allow to distinguish the ``confinement'' and the ``deconfinement'' phases is not known. In principle, any kind of phase transition should result in a non-analytic behavior of the free energy $\mathcal{F} \sim -\ln \mathcal{Z}$, which for the deconfinement transition in PCM is expected to behave as \cite{Cherman:14:1}
\begin{eqnarray}
\label{eq:free_energy_order_parameter}
\lim\limits_{N \rightarrow \infty}\mathcal{F}/N^2 \sim
\begin{cases}
1, & L_0 \Lambda \ll 1 , \\
0  & L_0 \Lambda \gg 1 .
\end{cases}
\end{eqnarray}
While a direct calculation of the free energy is nontrivial in Monte-Carlo simulations, first-order transitions typically result in a characteristic double-peak structure of the action probability distribution with unequal peak heights \cite{Bhanot:92:1}. Our numerical data for the mean energy does not exhibit any double-peak structure, which disfavours the first-order phase transition scenario (although not excluding it completely, as one might need very high statistics to distinguish the two peaks).

In the context of large-$N$ volume independence our results suggest that correlation length, mean energy and specific heat do not depend on the lattice size as long as it is much larger than the typical correlation length, in close analogy with large-$N$ gauge theories \cite{Neuberger:03:1,Neuberger:03:2}. The deviations from volume independence only become significant in the vicinity of the transition point or crossover.

\section{Twisted boundary conditions and the transition to \"{U}nsal-Dunne regime}
\label{sec:tbc}

We start the discussion of the principal chiral model with twisted boundary conditions (\ref{eq:boundary_conditions}) by presenting our results for the static correlation length (\ref{eq:correlation_length_Rossi}) on Fig.~\ref{fig:correlation_length}(b). If we plot $\xi$ as a function of $L_0$, we again clearly see two distinct regions separated by a peak of $\xi$: when $L_0$ is large the correlation length $\xi(L_0)$ coincides with zero-temperature value $\xi_0$ (\ref{eq:observable_0}). At intermediate values of $L_0$ $\xi$ exhibits a statistically significant growth, and in the region of small $L_0$ it decreases again, finally reaching some finite value. Note that the maximal relative change of $\xi$ is of the same order as in the case of periodic BC: approximately $5\%$. However, apart from the existence of the peak, the dependence of correlation length on $L_0$ and $N$ seems to be very different from the case of periodic BC. In particular, the peak height does not depend on $N$ within statistical errors, and its position shifts to smaller $L_0$ as $N$ is increased. The fits of the form (\ref{eq:observable_N_fit}) have rather poor quality for data points with fixed $L_0$.

Let us now recall that due to the properties (\ref{eq:twist_prescription}) of the twist matrix the group-invariant correlation function are periodic on the cylinder $\mathbb{R}\times S^1$ with effective size of $S^1$
\begin{eqnarray}
\label{eq:twisted_temperature}
\rho \equiv N L_0 .
\end{eqnarray}
In other words, the twist effectively increases the volume accessible to the system by a factor of $N$ and lowers the ``temperature'', which is a pre-requisite for the twisted Eguchi-Kawai reduction \cite{Unsal:08:1}. This property suggests that one should compare the data for twisted and periodic boundary conditions by identifying the compactification lengths as $N L_0^{TBC} = L_0^{PBC}$.

Plotting the static correlation length as a function of $\rho$ on the right plot on the top of Fig.~\ref{fig:correlation_length}, we observe a rather clear collapse of data points with different $N$ towards a single curve, which features a peak at $\rho_c/\xi_0 \approx 9$. At $\rho > \rho_c$ we find good agreement with data for periodic BC at zero temperature. Infinite-$N$ extrapolations using the fits (\ref{eq:observable_N_fit}) yield only very minor corrections to this picture. Increasing the spatial lattice size to $L_1 = 200$ at $N = 18$, we do not observe a significant enhancement of the correlation length beyond statistical errors, see the right plot on Fig.~\ref{fig:corr_length_volume_scaling}. We note here that the statistical errors in the correlation length appear to be larger for the twisted case, although we have almost two times more data points in this case. Unfortunately, large statistical errors do not allow us to make a definite conclusion on whether the enhancement of correlation length at $L_1 = 200$ is larger for periodic or for twisted boundary conditions. We can only say that for twisted case the change in the correlation length cannot be much larger than for periodic case.

\begin{figure}[h!tpb]
\centering
 \includegraphics[scale=0.8]{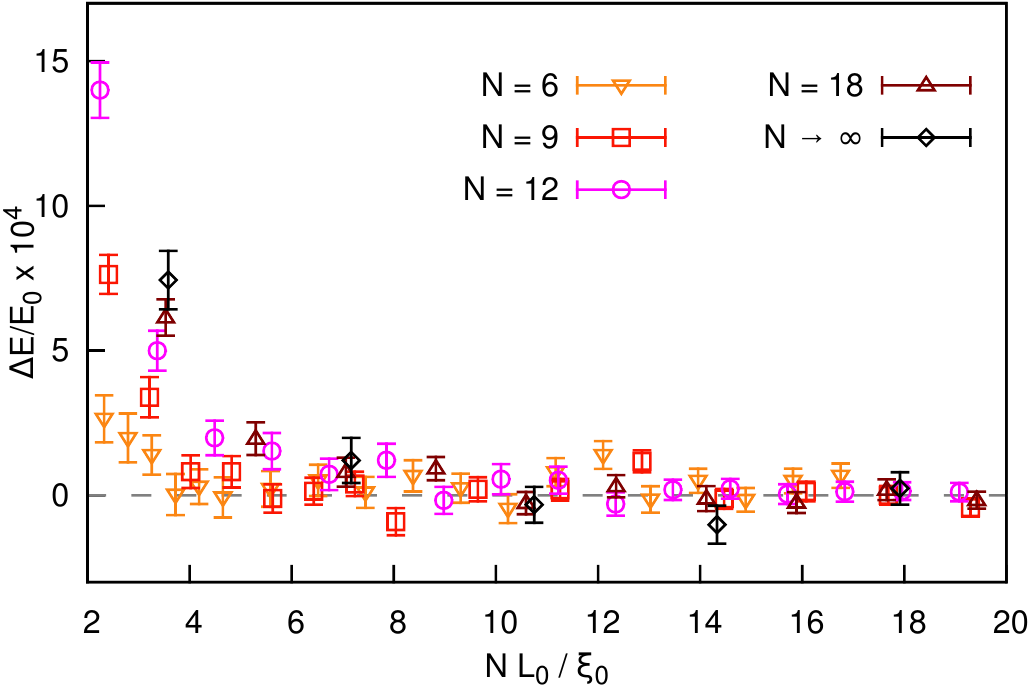}
 \caption{Relative change of mean energy $\Delta E/ E_0$ for twisted boundary conditions and different values of $N$ represented as a function of $\rho \equiv N L_0$.}
\label{fig:mean_energy_twisted_n}
\end{figure}

As one can see from Fig.~\ref{fig:mean_energy_and_specific_heat}, with twisted boundary conditions the mean energy and the specific heat practically do not depend on the compactification length down to the values of $L_0$ which roughly correspond to the position of the peak on Fig.~\ref{fig:correlation_length}(b). In contrast to the case of periodic boundary conditions, at small $L_0$ the mean energy $E(L_0)$ increases. The data for specific heat $C(L_0)$ has quite large statistical errors which probably do not allow to see non-trivial behavior on lattices $L_0 > 1$. Nevertheless, with smallest compactification length $L_0 = 1$ we again observe an increase of $C$ compared to zero temperature.

If we fix $L_0$ and fit the $N$ dependence of $\Delta E/E_0$ using (\ref{eq:observable_N_fit}), it extrapolates to zero in the large-$N$ limit for all values of $L_0$ which we consider ($L_0 \geq 2$). This observation supports the volume independence property at $\rho > \rho_c$. On Fig.~\ref{fig:mean_energy_twisted_n} we also plot the mean energy as a function of $\rho$. In agreement with volume independence property, for $\rho > \rho_c$ the values of $\Delta E/E_0$ are consistent with zero within statistical errors. However, in contrast to correlation length, for smaller values of $\rho$ different data points for the mean energy do not collapse. This suggests that for small values of $\rho$ long-distance quantities such as correlation length and local quantities such as mean energy might exhibit different scaling with $N$.

To summarize, with twisted boundary conditions we also observe some signatures of a nontrivial transition between the regimes of small and large compactification lengths which is controlled by a combined parameter $\rho \equiv N L_0$. Considered as a function of the compactification length $L_0$, this transition shifts to smaller and smaller $L_0$ as $N$ is increased (see the left plot in the bottom of Fig.~\ref{fig:correlation_length}), eventually approaching the zero radius limit at $N \rightarrow \infty$ and thus effectively disappearing. A similar behavior was found for the scale of dynamical symmetry breaking in gauge theories with unbroken center symmetry \cite{Unsal:08:2}. At $\rho > \rho_c$ physical observables practically do not depend on the lattice size, as could be expected for twisted Eguchi-Kawai reduction. The most important differences with the finite-temperature transition considered in the previous Section~\ref{sec:pbc} are, first, the independence of the height of the peak in the correlation length on $N$, and, second, the growth of the mean energy at small $\rho$, along with its nontrivial scaling with $N$. We will discuss these differences in more details in the concluding Section~\ref{sec:conclusions}.

\section{Non-perturbative saddle points}
\label{sec:gradient_flow}

Non-perturbative saddle points of the action in the path integral are one of the cornerstones of the physical applications of resurgence theory. In particular, factorial divergences in perturbative series which characterize small field fluctuations around non-perturbative saddles cancel similar divergences in perturbative expansion around the trivial vacuum saddle, thus allowing to complete the so-called resurgent triangle. To ensure that resurgent trans-series of the twisted compactified PCM in the \"{U}nsal-Dunne regime can be analytically continued to the low-temperature strongly coupled regime, it is also important to understand how the saddle points which dominate the path integral and enter the resurgent triangle change in the process of compactification from $L_0 \rightarrow \infty$ to $L_0 \ll 2 \pi \lr{\Lambda N}^{-1}$.

To study the features of the dominant saddle points in the path integral, we select randomly the field configurations generated in Monte-Carlo simulations, and evolve them along the path of the steepest descent towards one of the saddle points in its vicinity using the Gradient Flow equations \cite{Luscher:10:1}:
\begin{eqnarray}
\label{eq:gradient_flow}
 \frac{\partial U\lr{\xv, \tau}}{\partial \tau} = -\frac{i}{\beta N} \, \nabla^a_{\xv} S\lrs{U\lr{\xv, \tau}} \, T_a \, U\lr{\xv, \tau}, \nonumber \\
U\lr{\xv, \tau = 0} \equiv U(\xv),
\end{eqnarray}
where $\tau$ is the flow time and $\nabla^a_{\xv}$ is the $SU\lr{N}$ Lie derivative with respect to $U\lr{\xv}$:
\begin{eqnarray}
\label{eq:su_n_deriv}
\nabla^a_{\xv} f\lrs{U\lr{\yv, \tau} } = \left. \frac{d}{d s} f\lrs{e^{i s T_a \delta_{\xv,\yv}} U\lr{\yv, \tau}} \right|_{s \rightarrow 0} .
\end{eqnarray}
Here the $SU\lr{N}$ group generators $T^a$ are Hermitian and traceless matrices normalized as $\tr\lr{T_a T_b} = \delta_{a b}$. The advantage of using the Gradient Flow instead of other smoothing procedures such as smearing or cooling is that the Gradient Flow is continuous and reversible with respect to the flow time $\tau$, therefore it can be considered as a well defined change of the variables in the partition function which preserves all physical information encoded in the initial field configuration and at the same time ensures that in terms of the flow-evolved variables $U\lr{\xv, \tau}$ the partition function is dominated by smooth configurations \cite{Luscher:10:1}.

We have numerically solved equations (\ref{eq:gradient_flow}) using Runge--Kutta discretization scheme described in \cite{Luscher:10:1} with the time step $d\tau=0.1$ and initial conditions $U\lr{\xv, \tau=0}$ selected randomly from field configuration generated by Monte-Carlo process. We have continued the Gradient Flow up to the final flow time $\tau_f = 1.5 \times 10^3$.

In order to characterize the features of smoothed configurations we have considered the total action $S$ given by (\ref{eq:lattice_action}), as well as the local action density
\begin{eqnarray}
\label{eq:local_action_density}
S(\xv, \tau)
= \nonumber \\ =
\beta N \sum\limits_{i} \lr{N - \re \Tr\lrs{U\lr{\xv, \tau} U^\dagger\lr{\xv \pm \ev_i, \tau}}}
\end{eqnarray}
normalized such that it is zero for vacuum configuration with $U(\xv) = I$.

Since on a finite lattice the continuum saddle points such as unitons and fractons \cite{Unsal:14:1} are only approximate solutions to the saddle point equations, during the Gradient Flow evolution they appear as meta-stable states which are eventually destroyed. Nevertheless, within characteristic plateaus and not very large flow times the smoothed fields are expected to properly reflect the basic properties of the continuum saddle points (such as e.g. $Z_N$-valued holonomies or phases, or topology in gauge theories). For instance, point-like objects should appear in the profile of the actions density as pronounced lumps of action density on the smooth background. For very large flow times and in a finite volume this correspondence between smoothed fields and continuum saddle points is lost since Gradient Flow can be considered as a diffusion process which strongly entangles all degrees of freedom and spreads them uniformly on the lattice.

The fracton and uniton saddles which appear in the path integral of continuum two-dimensional PCM do not have an intrinsic topological structure due to the fact that $\pi_2[SU(N)] = 0$, in contrast to QCD and non-linear $\mathbb{CP}^{N-1}$ model. The absence of topological charge makes lattice studies of these non-perturbative objects more difficult and leaves the local action density (\ref{eq:local_action_density}) as the only scalar field which can characterize them in a simple and universal way.

In order to characterize the localization of action density for saddle point solutions, we have used the inverse participation ratio (IPR):
\begin{eqnarray}
\label{eq:ipr}
\ipr\lr{\tau} &=& V \left\langle \frac{ \sum\limits_{\xv} \tilde S^2(\xv, \tau)}{\lr{\sum\limits_{\xv} \tilde S(\xv, \tau)}^2} \right \rangle, \\
\tilde S(\xv, \tau) &=& S(\xv, \tau) - \min_{\xv} S(\xv, \tau),
\end{eqnarray}
where $\tilde{S}(\xv, \tau)$ is the action with subtracted constant background, $V = L_0 L_1$ is the lattice volume and averaging over smoothed field configurations at the same flow time is implied. By construction, this quantity takes the maximal value $\ipr = V$ when the action density is localized on a single lattice site and reaches the minimal value $\ipr = 1$ when it is everywhere constant. A very useful property of IPR is that it scales as $1/n$ if there are $n$ similar localized objects in the action density. In general, it gives the inverse fraction of the volume occupied by the support of $S(\xv, \tau)$, thus it can serve as a measure of action density localization.

\begin{figure}[h!tpb]
\centering
 \includegraphics[scale=0.82]{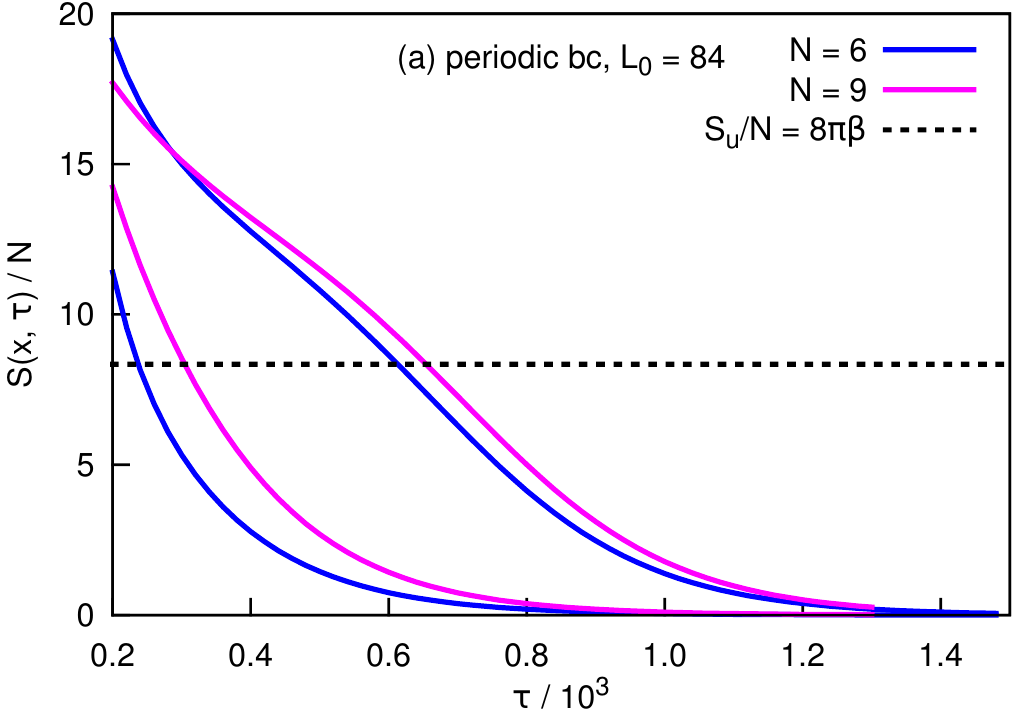}
 \caption{Dependence of the total action $S$ of smoothed configurations $U\lr{\xv, \tau}$ (\ref{eq:gradient_flow}) on the Gradient Flow time $\tau$ with periodic boundary conditions for $N = 6, \, 9$. Multiple solid lines represent independent Gradient Flows with different initial conditions $U\lr{\xv, \tau = 0} = U\lr{\xv}$ chosen randomly from field configurations generated by Monte-Carlo process. The black dashed line represents the continuum action $S_u = 8 \pi \beta N$ of the uniton. }
\label{fig:gradient_flow_actions_periodic}
\end{figure}

To present our results for periodic boundary conditions, on Fig.~\ref{fig:gradient_flow_actions_periodic} we first plot a typical dependence of the total action $S\lrs{U\lr{\xv, \tau}}$ of smoothed field configurations $U\lr{\xv, \tau}$ on the Gradient Flow time $\tau$, where several different lines represent independent Gradient Flows with different initial conditions $U\lr{\xv, \tau=0} \equiv U\lr{\xv}$. We observe that the total action starts from large values at initial moment of time and then rapidly decreases down to zero approximately at $\tau \approx 1.5 \times 10^3$. For some initial conditions, the decay of the action with the flow time becomes somewhat slower in the range $\tau = (0.4 \dots 0.8) \times 10^3$, so that sometimes a kind of ``plateau'' is formed.

\begin{figure*}[h!tpb]
\centering
 \includegraphics[scale=0.82]{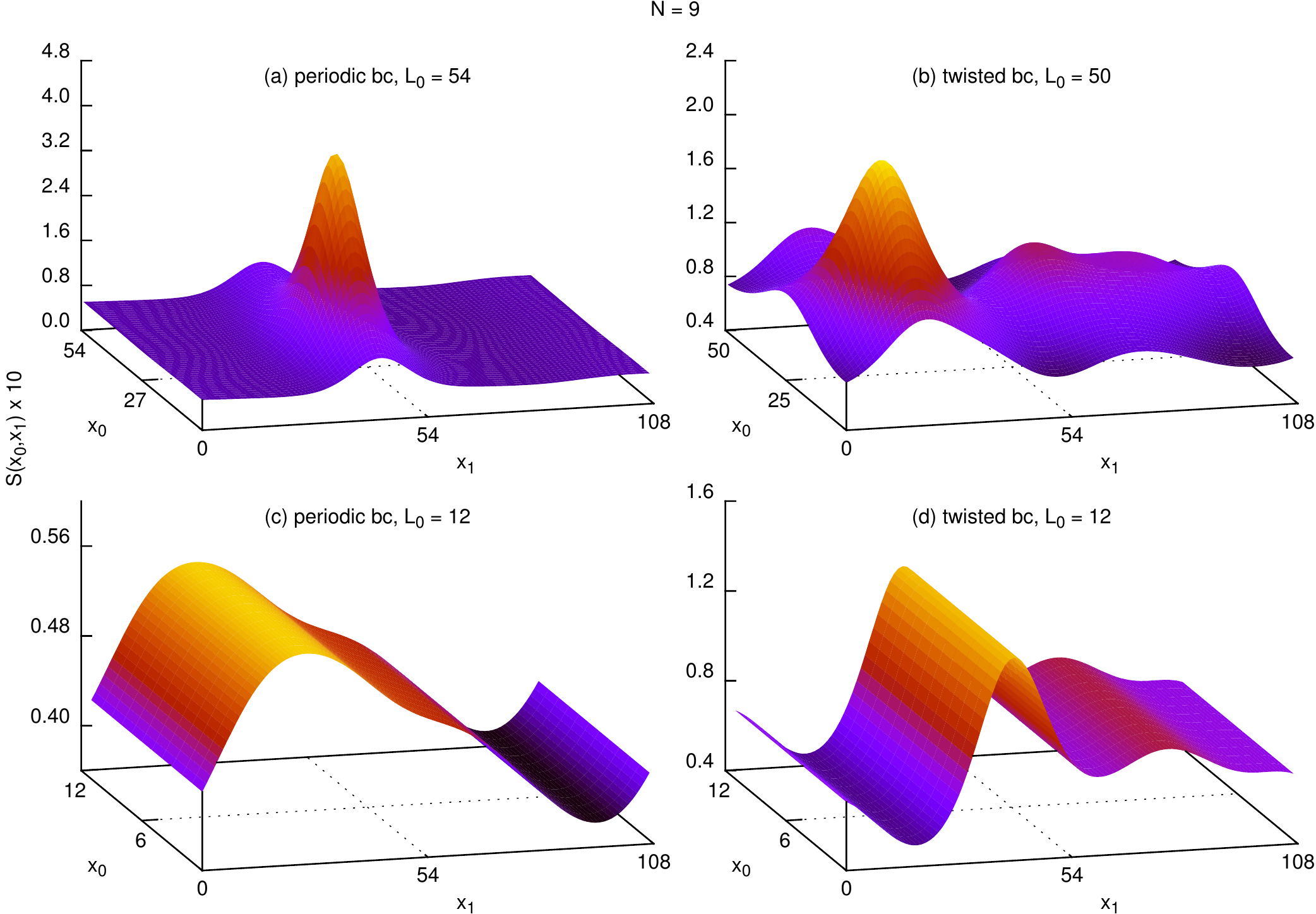}\\
 \caption{Typical action density $S(x_0,x_1)$ of smoothed field configurations $U\lr{\xv, \tau}$ (\ref{eq:gradient_flow}) taken at Gradient Flow time $\tau = 0.5 \times 10^3$ with periodic and twisted boundary conditions and different compactification length $L_0$ at $N=9$.}
\label{fig:saddles_N9}
\end{figure*}

Typical profiles of the action density of the smoothed fields taken at the characteristic ``plateau'' time $\tau = 0.5 \times 10^3$ are presented on the Fig.~\ref{fig:saddles_N9}(a) and  Fig.~\ref{fig:saddles_N9}(c) for large and small compactification lengths $L_0$, accordingly. For large compactification length $L_0$ the action density indicates the presence of some point-like objects which manifest themselves in pronounced action lumps with the size smaller than the length of the compact direction, whereas at compactification lengths smaller than the critical length as defined by the enhancement of static correlation length the saddle points become effectively flat along the compact direction.

\begin{figure}[h!tpb]
\centering
 \includegraphics[scale=0.82]{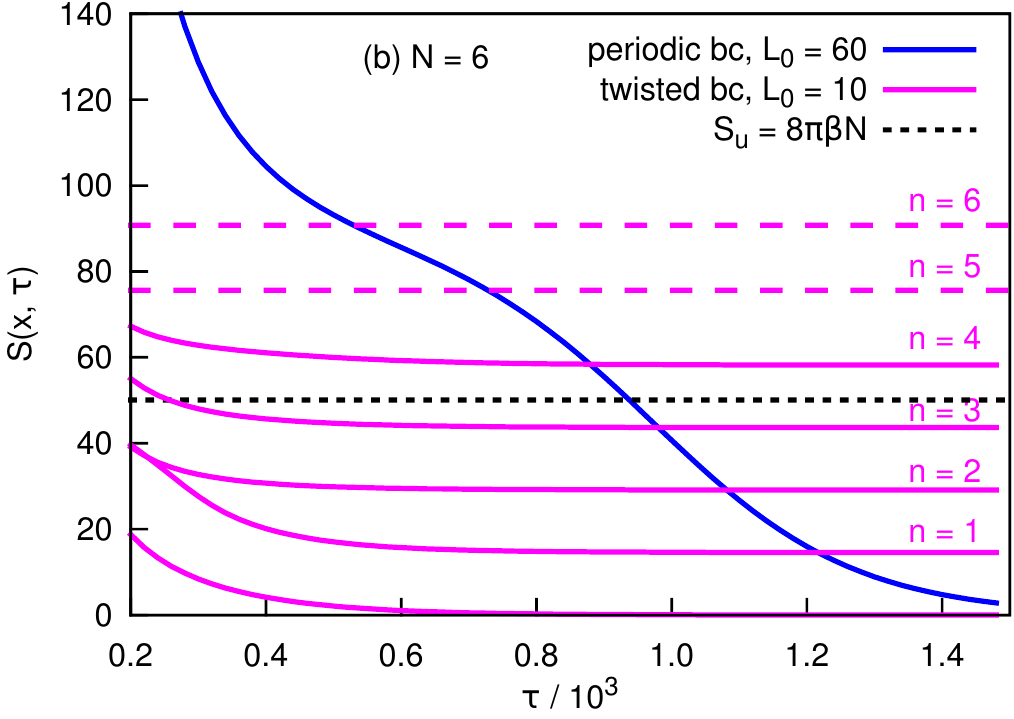}
 \caption{Comparison of the total action $S$ of smoothed configurations $U\lr{\xv, \tau}$ (\ref{eq:gradient_flow}) as a function of the Gradient Flow time $\tau$ for periodic and twisted boundary conditions at $N = 6$ and compactification length satisfying $N L^\mathrm{TBC}_0 = L^\mathrm{PBC}_0$. Multiple solid lines represent independent Gradient Flows with different initial conditions $U\lr{\xv, \tau = 0} = U\lr{\xv}$ chosen randomly from field configurations generated by Monte-Carlo process. The black dashed line represents the continuum action $S_u = 8 \pi \beta$ of the uniton. Long-dashed lines of magenta color represent extrapolated plateau values $S_p(n) = n S_p(n = 1)$ for twisted boundary conditions, which presumably correspond to fracton saddle points with the action $S_f = S_u/N$.}
\label{fig:gradient_flow_actions_twisted}
\end{figure}

\begin{figure*}[h!tpb]
\centering
 \includegraphics[scale=0.82]{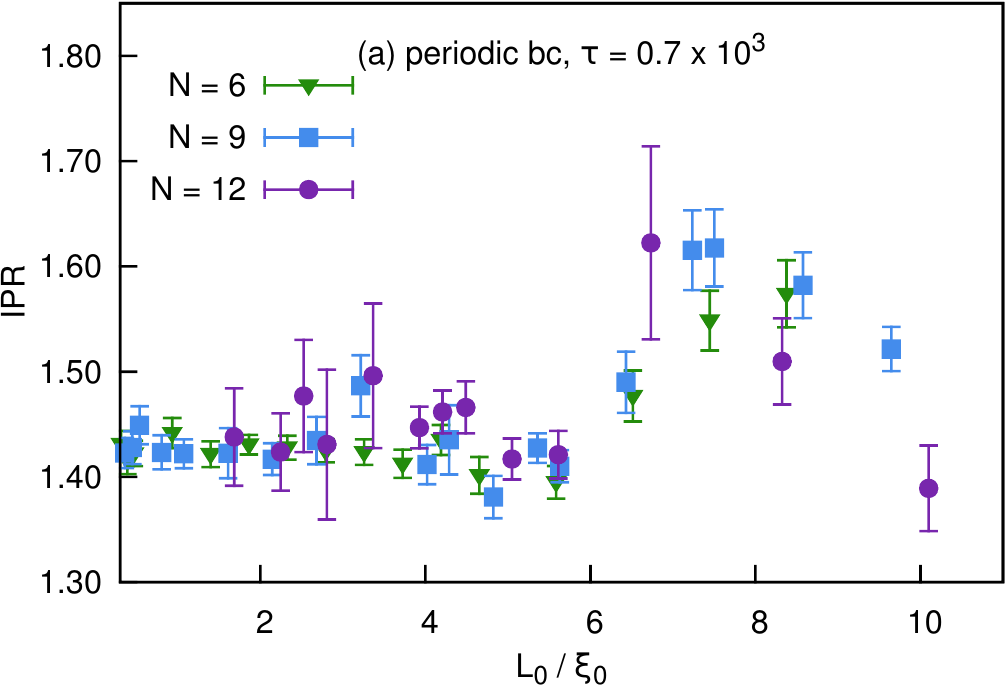}
 \includegraphics[scale=0.82]{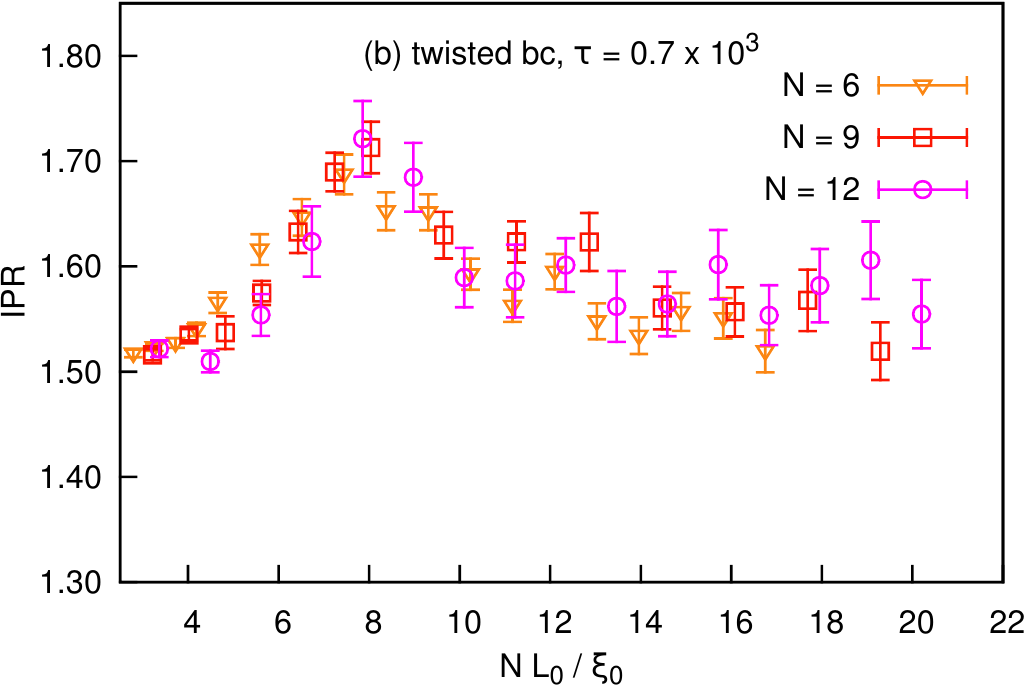}
 \caption{IPR (\ref{eq:ipr}) for smoothed field configurations at Gradient Flow time $\tau = 0.7 \times 10^3$ with periodic and twisted boundary conditions at different $N$.}
\label{fig:saddle_ipr}
\end{figure*}

Presumably, the particle-like objects which we observe in the large volumes can be associated with unitons, the well know unstable saddle points of the PCM. Unitons are harmonic maps $S^2 \rightarrow SU(N)$ \cite{Uhlenbeck:89:1}, where $S^2$ is obtained from $\mathbb{R}^2$ by including the point at infinity. Uniton solutions have the action which is quantized in units of
\begin{eqnarray}
\label{eq:uniton_action}
S_u = \frac{8\pi}{g^2} = 8 \pi \beta N
\end{eqnarray}
even in the absence of any well-defined topological charge. Within characteristic plateaus during the Gradient Flow evolution, the action is clearly proportional to $N$ (see Fig.~\ref{fig:gradient_flow_actions_periodic} for illustration), and agrees with $8 \pi \beta N$ ($8 \pi \beta = 8.34$ for our $\beta = 0.332$) within approximately $30\%$ uncertainty. In fact, one can't expect much better agreement due to non-zero contribution of ultraviolet fluctuations and renormalization of the coupling $\beta$ along the flow time. Here we do not consider the effect of this renormalization, since numerical extraction of renormalized $\beta$ would be quite complicated technically and is out of the scope of this work. The unstable nature of the uniton saddles qualitatively agrees with relatively short ``life-time'' of the non-perturbative objects which our Gradient Flow study reveals in the path integral of the PCM.

We further illustrate the dependence of the IPR (\ref{eq:ipr}) on $N$ and $L_0$ with periodic boundary conditions on Fig.~\ref{fig:saddle_ipr}(a). The Gradient Flow time is fixed to $\tau = 0.7 \times 10^3$. In the low temperature region the IPR features a rather wide peak, and takes larger values as compared to high temperature phase. The presence of the peak suggests that non-perturbative objects become more localized at these values of $L_0$. The maximum location strongly depends on $N$ and the flow time: we find that for larger flow times the peak moves towards larger $L_0$ with the shift being stronger for smaller $N$, which suggests that it is not directly related to the possible ``deconfinement'' transition discussed in Section~\ref{sec:pbc}.

Let us now turn to the Gradient Flow analysis of field configurations with twisted boundary conditions. We find that in large volume the picture is similar to the one for periodic boundary conditions. For \"Unsal-Dunne regime at small $L_0$ we present the total action of smoothed field configurations as a function of the Gradient Flow time on Fig.~\ref{fig:gradient_flow_actions_twisted} where we observe an important difference: there appear a number of well separated and very stable plateaus in the dependence of the action on the flow time, with very few transitions between them. With a rather good precision the action on these plateaus appears to linearly proportional to the plateau number, $S_p(n) = n S_p(n = 1)$, which hints at the emergence of effectively stable non-perturbative saddle points with quantized action. As for Monte-Carlo configurations the number of non-perturbative objects and hence plateaus which we observe is typically random, on Fig.~\ref{fig:gradient_flow_actions_twisted} we also plot some extrapolated plateau values for larger $n$.

The emergence of the new type of non-perturbative objects for twisted boundary conditions could be expected, since the twist introduces non-trivial potential on $SU(N)$ manifold effectively modifying it to the maximal torus $U(1)^{N-1}$ at low energies smaller that $1/(L_0 N)$. This potential has isolated minima with associated tunneling events between them which should appear as stable saddle points in Euclidean semi-classical description where stability is ensured by winding on the maximal torus \cite{Unsal:14:1,Cherman:14:1}. These winding numbers are responsible for the emergent topological structure and the stability of non-perturbative saddles in the \"Unsal-Dunne regime.

If we follow the twisted Eguchi-Kawai reduction prescription and identify compactification lengths with periodic and twisted boundary conditions as $L_0^{PBC} = N L_0^{TBC} \equiv \rho$, then from Fig.~\ref{fig:gradient_flow_actions_twisted} we find that lowest plateau action in twisted case is approximately $N$ times smaller than the uniton plateau action. Taking into account that uniton action is proportional to $N$, one can conclude that the lowest plateau actions is independent of $N$ with
\begin{eqnarray}
\label{eq:uniton_action_fractionalization}
 S_p^\mathrm{TBC}(N L_0) = S_p^\mathrm{PBC}(L_0) / N
\end{eqnarray}
at fixed $L_0$. This suggests the identification of these plateaus in the action with fracton saddle points, which are expected to carry the action
\begin{eqnarray}
\label{eq:fracton_action}
 S_f = S_u / N = 8 \pi \beta
\end{eqnarray}
and non-trivial Kaluza-Klein (KK) momentum $\xi_\mathrm{KK} = -2 \pi k /L_0$ \cite{Unsal:14:1,Cherman:14:1}. Counterparts of these stable fracton saddle points are well known in twisted $\mathbb{CP}^{N-1}$ model \cite{Bruckmann:09:1}: while in the large volume limit the details of boundary conditions should be irrelevant for instantons, with compactification length smaller than the size of instanton they split up to $N$ fracton constituents which carry fractional topological charge proportional to $1/N$. Solutions of $\mathbb{CP}^{N-1}$ equation of motions can be lifted to solutions of PCM equations of motion, therefore applying this procedure to $\mathbb{CP}^{N-1}$ instanton one yields in compactified \"Unsal-Dunne regime PCM unitons fractionalized into $N$ effectively stable constituents \cite{Unsal:14:1,Cherman:14:1}.

Typical action density profiles of non-perturbative objects by the Gradient Flow are given on Fig.~\ref{fig:saddles_N9}(b) and Fig.~\ref{fig:saddles_N9}(d) for large and small compactification length, correspondingly. As expected, we find that in the large volume non-perturbative objects are very similar to those with periodic boundary conditions. In contrast, saddles in the \"Unsal-Dunne regime with twisted boundary conditions are characterized by a much larger action than in the case of periodic boundary conditions with the same $L_0$, and are quite strongly localized.

However, we could not clearly observed the predicted fractionalization of unitons into $N$ fractons in the \"Unsal-Dunne regime. Rather, the maximal number of peaks which we have found in smoothed configurations have never exceeded $\sim 3$ regardless of $N$. As such, however, this is not a contradiction, since the number of peaks in the action density of smoothed configurations does not necessarily coincide with the number of fractional constituents of non-perturbative saddle points. With unimproved action, these constituents typically attract each other and eventually merge during the smoothing process \cite{Peschka:05:1}, sometimes before they become visible in the background of ultraviolet fluctuations.

The dependence of the IPR (\ref{eq:ipr}) of the non-perturbative saddle points with twisted boundary conditions is illustrated on Fig.~\ref{fig:saddle_ipr}(b). The IPR exhibits a rather sharp peak structure at intermediate values of $\rho = N L_0$, with the peak height being approximately independent of $N$. In contrast to periodic case, this peak moves approximately to the position of the peak in the static correlation length and at the same time becomes smaller and narrower as flow time becomes larger. This coincidence of peaks in IPR and the static correlation length might indicate some nontrivial rearrangement of non-perturbative objects in the process of transition to the Dunne-\"Unsal regime. We note, however, that even for the simplest model of the ideal gas of extended non-perturbative objects the IPR might exhibit non-monotonic behavior which is just the reflection of the competition of two scales - the lattice size and the characteristic size of non-perturbative objects. Thus while the sharp peak in the IPR on Fig.~\ref{fig:saddle_ipr}(b) might be an indication of some nontrivial transition in the structure of non-perturbative saddles, this indication should not be considered as conclusive.

\section{Conclusions}
\label{sec:conclusions}

In this paper we have studied possible signatures of a crossover or a phase transition between the regimes of small and large compactification lengths $L_0$ for the two-dimensional $SU\lr{N} \times SU\lr{N}$ principal chiral model (PCM) both with periodic and with twisted boundary conditions. By analogy with other asymptotically free field theories one expects some kind of ``deconfinement'' transition for periodic boundary conditions \cite{Cherman:14:1}. According to the adiabatic continuity conjecture, the twist is expected to eliminate this transition \cite{Unsal:14:1,Cherman:14:1,Sulejmanpasic:16:2}, so that the regimes of small and large compactification lengths can be analytically related.

In the absence of well-defined local order parameters such as e.g. Polyakov loop, we have considered universal physical observables which can characterize phase transitions regardless of the symmetries of the system: mean energy, specific heat and static correlation length. We have found that for both types of boundary conditions these quantities behave in a way which is compatible with the signatures of a rather weak crossover or a phase transition: mean energy and specific heat exhibit monotonic growth/decrease with compactification length once it is sufficiently small, and the static correlation length is enhanced near some ``critical'' compactification length.

An important difference between the two boundary conditions is that for periodic boundary conditions the peak in the correlation length becomes somewhat higher and narrower as the $SU\lr{N}$ rank $N$ is increased. It also becomes slightly higher for larger lattice volumes. Since the large-$N$ limit can be also considered as thermodynamic limit within the range of validity of Eguchi-Kawai reduction, this behavior suggests that PCM with periodic boundary conditions might indeed feature a finite-temperature phase transition, at least in the large-$N$ limit.

In contrast, for twisted boundary conditions the shape of the peak in the static correlation length is independent of $N$, once the data is considered as a function of the combined length parameter $\rho = N L_0$. The dependence of the peak height on the spatial lattice size also cannot be distinguished within statistical errors. This behaviour is not typical for a phase transition, but can be still compatible with a weak crossover. If true, the crossover scenario would be a challenge for the continuity conjecture, since the phases separated by the crossover typically cannot be analytically related to each other (a classical example is the Berezinskii--Kosterlitz--Thouless transition).

By using the Gradient Flow, we have also studied the structure of non-perturbative saddle points which dominate the path integral of the PCM with both boundary conditions. We have found localized non-perturbative object with the properties expected for unitons, the unstable saddle points of the continuum PCM. In particular, these objects have quantized action which scales linearly with $N$. As expected, they also become effectively stable for twisted boundary conditions, thus exhibiting the phenomenon of emergent topology \cite{Unsal:14:1}. We also find that for twisted boundary conditions the geometric properties of non-perturbative saddles change precisely at the position of the possible crossover to the Dunne-\"Unsal regime, which is yet another argument that this crossover might be non-trivial.

\begin{acknowledgments}
This work was supported by the S.~Kowalevskaja award from Alexander von Humboldt Foundation. We thank F.~Bruckmann, A.~Cherman, A.~Dromard, G.~Dunne, P.~Orland, T.~Sulejmanpasic, M.~Unsal and A.~Zhitnitsky for useful discussions. We are also indebted to A.~Cherman, G.~Dunne, T.~Sulejmanpasic and M.~Unsal for helpful comments on this manuscript.
\end{acknowledgments}


%





\end{document}